\documentstyle[epsf]{article}

\newcommand{\be}{\begin{equation}}
\newcommand{\ee}{\end{equation}}

\begin{document}

\begin{flushright}
Liverpool Preprint: LTH 531\\
 \end{flushright}
  
\vspace{5mm}
\begin{center}
{\LARGE \bf Hybrid meson decay from the lattice}\\[10mm] 
{\large\it UKQCD Collaboration}\\[3mm]
 
 {\bf   C. McNeile,   C.~Michael\\
Theoretical Physics Division, Dept. of Mathematical Sciences, 
          University of Liverpool, Liverpool L69 3BX, UK \\
 and P. Pennanen\\ 
Department of Physical Sciences, Theory Division,
 University of Helsinki,
FIN-00014 Finland  }\\[2mm]

\end{center}

\begin{abstract}

 We discuss the allowed decays of a hybrid meson in the heavy quark 
limit. We deduce that an important decay will be   into a heavy
quark non-hybrid state and a light quark meson, in other words, the
de-excitation  of an excited gluonic string by emission of a light
quark-antiquark  pair. 
 We discuss the study of hadronic decays from the lattice in the heavy
quark limit and apply  this approach to explore the transitions from a
spin-exotic hybrid  to $\chi_b \eta$ and $\chi_b S$ where $S$ is a
scalar meson.  We obtain a signal for the transition emitting a scalar
meson and we discuss  the phenomenological implications. 

\end{abstract}

\section{Introduction}

 Hybrid mesons are those with non-trivial excited gluonic components.
The simplest  such case is when the spin-parity is exotic, namely not
allowed in the quark model. Here we specialise to heavy quarks and so 
our comparisons with experiment will be for $b \bar{b}$ systems. In this
 context, there will be a spin-exotic ($J^{PC}=1^{-+}$) meson whose 
properties can be determined from lattice QCD. We review here first the 
information on the nature and spectrum of such excited gluonic states.
We  then discuss in general the allowed decay modes of such a state. In
most of this  discussion we focus on predictions in the heavy quark
limit, so with  heavy quark spin-flip neglected.

 We then review lattice methods to extract hadronic transition matrix
elements.  In the case of hybrid decay, we explore the creation of a
light quark-antiquark state  from the gluonic field of the hybrid meson.
It is possible to fulfil the rather restricted conditions on a lattice
and we are able to explore  these transitions.  We study hybrid meson
transitions  to $\chi_b \eta$ and $\chi_b S$ where $S$ is a scalar
meson.  We obtain a signal for the transition emitting a scalar meson
and we discuss  the phenomenological implications.

\section{Hybrid states on the lattice}

The static quark approach gives a very straightforward way to explore
hybrid quarkonia.  These will be $Q \bar{Q}$ states in which the gluonic
contribution is excited.  The ground state of the gluonic degrees of
freedom has been explored on the lattice,  and, as expected, corresponds
to a symmetric cigar-like  distribution  of colour flux between the two
heavy quarks at separation $R$. One can then construct less symmetric
colour distributions which would correspond to gluonic excitations. For
a review see ref.~\cite{hf8}.  The properties of the physical states can
then be obtained from these static potentials by solving the
Schr\"odinger equation in the adiabatic approximation.

The way to organise this is to classify the gluonic fields according to
the symmetries of the system.  This discussion is very similar to the
description of electron wave functions in  diatomic molecules. The
symmetries are  (i) rotation around the separation axis $z$ with
representations labelled by $J_z$ (ii) CP with representations labelled
by $g$ and $u$ and (iii) C$\cal{R}$. Here  C interchanges $Q$ and
$\bar{Q}$, P is parity and $\cal{R}$ is a rotation  of $180^0$ about the
mid-point around the $y$ axis. The C$\cal{R}$ operation is only relevant 
to classify states with $J_z=0$. The convention is to label states of
$J_z=0,1,2$ by $ \Sigma, \Pi, \Delta$  respectively. 

\begin{figure}[ht] 
 \epsfxsize=10cm\epsfbox{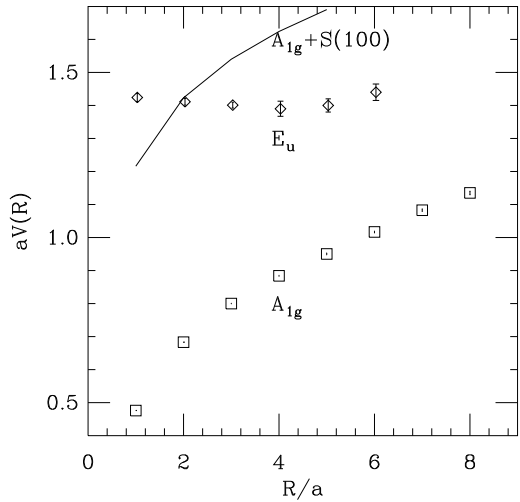}

  \caption{The ground state ($A_{1g}$) of the  static  potential $V(R)$
and  the first gluonic excitation ($E_u$) from this work with $N_f=2$
flavours  of sea quark (of approximately the mass of the strange quark),
 in lattice units with $a \approx 0.1 $fm. The energy of a scalar  
meson with momentum $ \pi/8a$ above the ground state potential is shown 
by the continuous line.
 }
   \label{pmvr}
  \end{figure}

In lattice studies the rotation around the separation axis  is replaced
by a four-fold discrete symmetry and states are labelled  by
representations of the discrete group $D_{4h}$.  The ground state
configuration of the colour  flux is then $\Sigma^+_g$ ($A_{1g}$ on the
lattice). The exploration of the energy levels  of other representations
has a long history in lattice studies~\cite{livhyb}. The first excited
state is found  to be the $\Pi_u$ ($E_u$ on a lattice) - see
figure~\ref{pmvr}  for an illustration. This can be visualised  as the
symmetry of a string bowed out in the $x$ direction minus the same 
deflection in the $-x$ direction (plus another component of  the
two-dimensional representation with the transverse direction $x$
replaced by $y$), corresponding to flux  states from a lattice  operator
which is the difference of U-shaped paths from quark to antiquark of the
form $\, \sqcap - \sqcup$.

 A summary of lattice determinations of the energy of this lowest hybrid
state~\cite{hf8}  puts it at $m(H)=10.76(7)$ GeV for $b$ quarks, so
approximately 1.3 GeV  heavier than the $\Upsilon$. This hybrid state 
in the adiabatic approximation will have lowest angular momentum $L=1$
and combining this with the heavy quark spins gives 8 degenerate $J^{PC}$ 
values. Of special interest is the spin exotic state with $J^{PC}=1^{-+}$ 
which is expected to be the lightest spin-exotic meson. Since it 
is spin-exotic, it cannot mix with the non-hybrid $Q \bar{Q}$ states and is 
thus of considerable theoretical and experimental interest.

\section{Hybrid meson decays}

 We shall be discussing hybrid meson decays in the heavy quark limit -
so  our conclusions will be more applicable to $b$-quark systems than
$c$-quarks. From the adiabatic approximation, one solves the gluonic
field around a  static quark-antiquark at separation $R$ to determine a
potential first and then solves the Schr\"odinger equation in that
potential.  For a spin-exotic hybrid with an excited gluonic field
having $J_z=1$ about the interquark axis, it follows that the
quark-antiquark  must have orbital angular momentum greater than or
equal to one unit  and we will assume the least angular momentum
allowed, namely a P-wave. Hence the  quark-antiquark system has $L=1$ in
the hybrid and this will persist to the final state  in any decay.
Furthermore the spin-exotic hybrid has the heavy quark-antiquark in a
spin triplet so this (and its spin projection ) will also persist to the
final state in any decay. The potential that binds the hybrid meson is
relatively flat (see fig.~\ref{pmvr}). 
 Hence the hybrid meson has an   extended radial wavefunction, for
example~\cite{jkm} with $R$-dependence of  $u_H \approx R^2
e^{-(R/0.51)^2}$ in units of fm for the inter-quark coordinate $R$ for
$b$ quarks. 
 This has implications for their production and decay. For instance, 
any vector hybrid state will only be weakly produced in $e^+ e^-$
collisions because the wave  function at the origin is  suppressed.

 We shall later consider transitions at fixed $R$ from the hybrid state
to the  P-wave $\chi_b$ state with radial wavefunction approximately
$u_{\chi} \approx R^2 e^{-(R/0.33)^2}$ with $R$ in fm. The lack of nodes
in the relevant wave functions implies  a spatial wave function overlap
factor which is quite large (ie  assuming the transition rate is 
independent of $R$, for the above normalised wavefunctions $\int u_H
u_{\chi} dR = 0.63$).

 Given the mass estimate above, the open channels for decay of a 
$J^{PC}=1^{-+}$ hybrid include  $ B\bar{B^*},\ B^* \bar{B^*},\ \eta_b
\eta,\ \eta_b \eta',\ \chi_{b} \eta,\ \chi_b S,\ \Upsilon(1s) \omega$
and $ \Upsilon(1s) \phi$  where $S$ is a scalar meson which can
subsequently decay to $\pi \pi$ (note decay to $B\bar{B}$ is not allowed
by $C$ conservation). However, as discussed above,  decays to  
quarkonia are only allowed into a $\chi_b$ state in the heavy quark
limit. Thus  decays to $\eta_b$ or $\Upsilon(1s)$  proceed by  heavy
quark symmetry violations (of order $1/M_b$). We do not discuss these
modes further here.

  Selection rules have been proposed for hybrid decays, for
example~\cite{page} that $H \not \to X + Y $ if $X$ and $Y$ have the
same non-relativistic structure and each  has $L=0$. This would rule out
$B\bar{B}$, $B\bar{B^*}$  and $B^* \bar{B^*}$ and the analogous cases for
charm quarks. This selection rule can be addressed directly from the
static quark  approach. The symmetries in this case of rotations and
reflections about the separation axis have to  be preserved in the
strong decay. From the initial state with the  gluonic field in a given
symmetry representation, the $q \bar{q}$ pair must be produced in   the
decay in such a way that the  combined symmetry of the quark pair and
the final gluonic distribution matches  the initial representation.

We first discuss decays of non-hybrid quarkonia to set the scene.  For
the ground state of the gluonic excitation  ($\Sigma^+_g$, non-hybrid)
we have $J_z=0$ and even $CP$. Thus, for this state to decay to
$(Q\bar{q})(\bar{Q}q)$ with each heavy-light meson having $L=0$, the
final gluonic  distribution is also spatially symmetric about the
separation axis (actually  it is essentially two spherical blobs around
each static source binding the  heavy light mesons). Then any $q
\bar{q}$ pair production has to respect this symmetry and have $J_z=0$
and even $CP$. Since each light quark has no orbital angular momentum
about the separation axis, the $CP$ condition then requires $S_{q
\bar{q}}=1$, a triplet state.  This conclusion for the light quark spin
assignment can be tested by the ratio of  $B\bar{B}, B \bar{B^*} $ and
$B^*\bar{B^*}$ decays.

We now consider  decays from  a heavier quarkonium state to a lighter
such state (with both intial and final quarkonia in the most-symmetric
gluonic $\Sigma_g^+$ representation) with   emission of a light
quark-antiquark pair which form a flavour-singlet meson. This flavour
singlet meson must be produced with $CP$ even and  $J_z=0$ again.
Possible modes are a scalar meson  or a vector meson with $S_z=0$ which
are both allowed in a symmetric spatial state. Note that decay to  a
pseudoscalar meson is not allowed since the required spatial 
wavefunction would have to be  in a $\Sigma^+_u$ representation but this
is not realisable for a meson with no spin. 

  The spin nature of the  quark-antiquark pair produced in hadronic
decays has been widely discussed~\cite{3p0}. 
  A colour-singlet light quark-antiquark pair can be produced from
colour flux oriented in the $z$-direction  with vacuum quantum numbers
($CP=+1,\ J_z=0$, flavour singlet) either as a scalar meson ($^3 P_0$
model)  or as the zero-helicity component of a vector meson ($^3 S_1$
model). The lattice study of flavour singlet mesons suggests that  the
former is a much larger~\cite{ozi} amplitude in general. In any specific
case, however,  the amplitudes can be determined explicitly, and this we
undertake for hybrid decays.

 For the $J^{PC}=1^{-+}$ hybrid  we have a gluonic field with $J_z=1$
and odd $CP$.  For the case of decay to a $(Q\bar{q})(\bar{Q}q)$ with
each heavy-light meson having $L=0$, this would imply that the 
$q\bar{q}$ would have to be  produced with $J_z=1$ and odd $CP$. This is
not possible since the triplet state  would have even $CP$  while the
singlet state cannot have $J_z=1$. This is then equivalent to the
selection rule  described above. There will presumably be small
corrections to this selection  rule coming from retardation effects.
 Decay to $(Q\bar{q})(\bar{Q}q)$ with one heavy-light  meson having a 
non-zero orbital excitation is allowed from symmetry but is not allowed
energetically with conventional mass assignments~\cite{bmass} for the
P-wave excited  B meson multiplet.

 Decays of a hybrid meson to $(Q\bar{Q})(q\bar{q})$ are also possible
since there is enough  excitation energy to create a light quark meson. 
This meson must be created in a flavour singlet state and  the lightest
candidates are $\eta$, $\omega$ and scalar ($S$) channels. In a lattice
context, this production is via a disconnected quark loop while the
normalisation of the meson  will involve the connected correlator. Thus
the relative strength of the disconnected  correlator to the connected
correlator enters and this has been studied for different meson  quantum
numbers on a lattice~\cite{ozi}. As expected from the phenomenology of
meson spectra, the  pseudoscalar and scalar mesons are the only two
cases with relatively  large disconnected contributions. For
pseudoscalar mesons,  the flavour singlet mixture of $\eta$ and $\eta'$ 
is mainly $\eta'$ with only an amplitude of $\sin(10^0)$  of $\eta$ for
the conventional mixing scheme (see~\cite{cmcmeta}).  For the scalar
meson, the discrete states coming from mixing of the glueball  and  $q
\bar{q}$ meson are relatively heavy so may not be allowed from energy
considerations,  but one should  also consider the $\pi \pi$ continuum
with favour singlet  and scalar quantum numbers, which is experimentally
known to be big (ie have large $\pi \pi$ phase shift) around 700 MeV.

 In these decays from a hybrid meson with $\Pi_u$ representation to a 
de-excited string with $\Sigma^+_g$ representation, the light quark
meson  must have a wavefunction with a net $\Pi_u$ representation with
$J_z=1$ and $CP=-1$. If this meson has an angular momentum of $L$ about
the  final state heavy-heavy meson, then this implies that the orbital
wave function  has $L_z \le L$ and $CP = (-1)^L$. Using this orbital 
angular momentum, then allows the  decay products from the decay of  a
$J^{PC}=1^{-+}$  meson to be identified.
 Thus in the static limit the  decays  allowed by symmetry  for a
$\Pi_u$ representation hybrid to a $\Sigma^+_g$ representation state
plus flavour-singlet meson are shown in Table.~\ref{hdecaytable}. The
examples shown take account of the adiabatic approximation and the nature
of the $J^{PC}=1^{-+}$ hybrid wavefunction.

  \begin{table}[h] 
 \begin{tabular}{ccccccl}
\hline
   Meson  &  $J^{PC}$   &  wave fn.& $L_z$ & $CP$ & $L$ &  Example\\
\hline
 $\eta$, $\eta'$ & $0^{-+}$ & $\Pi_g$ & 1 &+ &  2 & $ {\rm Hybrid} \to
\chi_b + \eta$\\
 scalar & $0^{++}$ & $\Pi_u$ & 1 &- &  1 & $ {\rm Hybrid} \to \chi_b
+\pi+\pi$\\
\hline

\end{tabular}
 \caption{Hybrid decays by string de-excitation in the heavy quark limit
 emitting a flavour singlet light quark-antiquark meson with quantum
numbers $J^{PC}$. This meson has a wavefunction relative to the heavy
quark-antiquark system  in the representation shown with $L_z$ and $CP$
as shown.  This implies that it is  in an  orbital $L$-wave about the
heavy quark system.}
 \label{hdecaytable}
\end{table}

 As for the case of quarkonium decays and string
breaking~\cite{cmpp,MILC}, it is possible in principle to explore on the
lattice some aspects of these hybrid meson decays. One can study matrix
elements between ground states which are degenerate in energy such as 
the $1^{-+}$ hybrid and the $\chi_{b} \eta$ final state where the light
quark  mass is adjusted so that there is equal energy in both systems. 
This and similar lattice studies will enable some further  guidance to
be given for experimental searches for hybrid mesons.

 \section{Decays from the lattice}

 Consider the generic transition $H \to A + B$ where $A$ and $B$
represent  stable particles and $H$ is unstable to decay. Here we assume
that the two-body state has exactly all the symmetries of the state
$H$. For  simplicity we will consider $H$ at rest and then $A$ and $B$
have momenta $k$ and $-k$ respectively. Thus the two-body state, if
non-interacting, has energy  $E_{AB}= \sqrt{m_A+k^2} + \sqrt{m_B+k^2}$. 
 In Euclidean time, the properties of this decay transition are very 
different in practice~\cite{cmdecay} from the Minkowski case, in
particular the large time correlator will be dominated by the lightest
two-body state which will be that with minimum momentum. 

 One way to explore this system in detail is to consider a finite
volume. We make the usual assumption that the theory is defined
independently of the  boundary conditions. From the lattice viewpoint,
the finite volume result can be obtained by taking the  continuum limit
at fixed physical volume.  For a cubic spatial volume $L^3$ with
periodic boundary conditions, the  momenta are discrete (${\bf k} = 2
\pi {\bf n}/L$) where  ${\bf n}=(n_1,\ n_2,\ n_3)$ is an integer vector.
The two-body  states are then also discrete in energy. One expects that
as $L$ increases  beyond the range of the two body interaction, the two
body energy  levels become close to the non-interacting case. This has 
been studied~\cite{lu91} and detailed formulae obtained for the energy
shifts at sufficiently large $L$ in terms of the scattering phase shift
in the $ A + B $ system, provided inelasticity is negligible. This
allows, in principle,  to measure the phase shift at various energies by
varying $L$ and ${\bf n}$. From the phase shift one can then deduce  the
properties of the decay in the large volume limit. To  measure the
lightest two-body state accurately (typically with $n=(1,\ 0,\ 0)$) is
already a challenge and to obtain  accurate energy determinations for
excited states with higher momentum will  be much harder. Moreover,  in
practice the energy shifts are small and so it will be extremely
difficult to measure  accurately the phase shifts on a
lattice~\cite{aoki}.

 For some applications, it is possible to measure the transition
amplitude  directly. This is clearly the case in a  quenched (or partially
quenched)  approach where the decay transition does not actually take
place in the  lattice version of the theory. For example, the $\rho$
meson does not  decay to $\pi + \pi$ in quenched studies. Let us
describe how this can be measured  in principle:
 Create $H$ at $t=0$ and annihilate a two-body state with relative
momenta $k$ and $-k$ at  time $t$.
 Then the  contribution to the correlator from a $H$ state with mass $m_H$ 
and a two-body state with energy $E_{AB}$ is given by
 \be
   C_{H-AB} (t)=\sum_{t_1} h e^{-m_H t_1} x e^{ -E_{AB} (t-t_1)} b
 \label{3pt}
 \ee
 where the summation over the intermediate $t$-value  $t_1$ will be an
integral in the continuum and  where $h$ and $b$ are the amplitudes to
make each state from the lattice operators used  and $x$ is the required
transition amplitude $\langle H | AB \rangle$.  Here we are assuming
that the states $H$ and $AB$  are normalised to 1.  By obtaining $h$ and
$b$  from the $H \to H$ and $AB \to AB$ correlators,  one can
hope~\cite{cmpp,cmcm} to  extract $x$. 

 The complication, however, is that removal of excited state
contributions is  tricky. For example, if $m_H - E_{AB}  > 0 $ then
the transition time $t_1$  will be preferentially near 0 (since the
heavier state then propagates less  far in time) and one can complete
the sum over $t_1$ obtaining a  $t$-dependence of eq.~\ref{3pt} as
$e^{-E_{AB}t}$. This same $t$-dependence  would be obtained if the state
with mass $m_H$ were to be replaced with an  excited state with an even
heavier mass. Thus one cannot separate the  ground state and excited
state contributions even in principle. See ref.~\cite{cmcm} for 
a fuller discussion.

 The way forward is that if $m_H = E_{AB}$, the ground state
contributions have a $t$-dependence  as $t e^{-E_{AB}t}$ whereas any
excited state contributions behave as $e^{-E_{AB}t}$  as above. So now
we do have a way to isolate the required ground state  contribution:
 \be
    x = \lim_{t \to \infty}
  { 1 \over t} {C_{H-AB} (t) \over  [C_{H-H} (t)\ C_{AB-AB} (t)]^{1/2} }
 \label{xrat}
 \ee

Note that this separation is only by a power of $t$ which is  less than
the case for diagonal  correlations where the excited state 
contributions are suppressed by an exponential $e^{(m'_H-m_H)t}$.  
 
  In practice the requirement of energy equality can be relaxed.
Defining $\Delta=  m_H - E_{AB}$, then the ground state contribution
to the expression of eq.~\ref{xrat} evaluates  to   $2 x \sinh
(t\Delta  /2)/(t\Delta ) = x(1+(t\Delta )^2/24 + \dots)$. So this will be
equivalent to the expression  with $\Delta=0$ provided
 \be
   (m_H - E_{AB}) t << 5
 \label{onshell}
 \ee

\begin{figure}[h]

\begin{center}
\epsfxsize=8cm\epsfbox{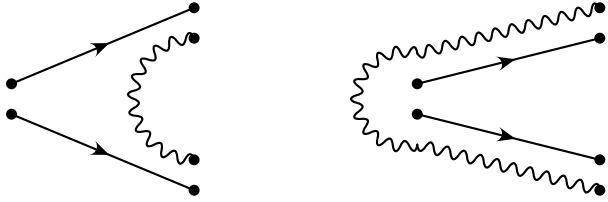}

 \caption{ Light quark pair production (wiggly lines) for the three
point function $H \to A + B$  in Euclidean time (running horizontally).
The straight lines represent quarks which may be heavy or light. The left
hand diagram has  the interpretation of a transition (our $x$) at an
intermediate  time while the right hand diagram can be thought of as 
some intrinsic mixing in the $H$ state.
 }
 \label{3ptfig}

 \end{center}
\end{figure}

 So far we have described the behaviour of the  $C_{H-AB} (t)$ in the
quenched  approximation. In full QCD, there will be a  mixing of these
two states.  Let us illustrate this for the case of interest where the
energies are  approximately the same (namely $E$). Then the energy
mixing matrix has the form
 \be
 \left( \matrix{ E & x \cr
                 x & E }   \right)
 \label{energy}
 \ee
 which has eigenvalues $E \pm x$. An accurate measurement of these
energy  eigenstates would then give the transition amplitude $x$. If $x$
is numerically  small, it is actually possible to follow an approach
similar to that described above for the quenched  approximation. Namely,
if $x$ is small, one can work to a given low order in $x$.  Then
provided eq.~\ref{onshell} is satisfied, to first order in $x$, we again
find that  $ C_{H-AB} (t)$ will have a contribution with a
$t$-dependence behaving as $xte^{-E_{AB}t}$ from eq.~\ref{3pt}  just as
described above. As well as further  transitions and corrections from
the mixing energy shifts which will both be of higher order in $xt$, one
must also consider the intrinsic mixing of the initial $H$ state with
$AB$ (and vice versa). This intrinsic mixing (ie not the mixing induced
by the propagation  from the energy matrix of eq.~\ref{energy} - see
fig.~\ref{3ptfig} for an illustration of a  typical contribution) is
expected be of order $x/{\cal E}$ where ${\cal E}$ is the energy of the
quark pair and so  will contribute a term like $x e^{-E t}/{\cal E}$.
This is a contribution  similar to that from excited states  and so will
 be dominated at large $t$ by the $xt e^{-E t}$ term we are looking for.
So we need both $xt$ to be small and $t$ to be large. This implies that 
$x$ must be small for this simplified approach.

 We now discuss whether $x$ is generically small. Provided the ranges of
the  interactions between $A$ and $B$ and between $H$ and $AB$ are
effectively  finite and smaller than the spatial extent $L$, then the
transition probability  $x^2$ will be proportional to $1/L^3$ and hence
the transition amplitude  $x$ behaves as $1/L^{3/2}$. As $L$ is
increased, the different momentum  states of $A+B$ become closer
together in energy and the density  of states behaves like $L^3$. Hence
the net transition probability to  states close to a given momentum will
be independent of $L$ at large $L$  as expected. Thus we conclude that
$x$ is indeed small at large volume but that off diagonal transitions
between different momentum states  will become  important.  Thus at
large volume there will be many small $x$'s to take into  account.

 So a practical method will be possible if the lowest energy $AB$ state 
that couples to $H$ has a similar energy to $H$. This lowest energy
state  will have relative momentum ${\bf n}=0$ for S-wave decays and 
${\bf n}=(1,\ 0,\ 0)$ for P-wave decays etc.  By adjusting the  lattice
volume and quark mass, it may be possible to arrange for  approximate
energy equality: this is often called an on-shell transition. From
studying the
 correlations as above, one can then extract the transition amplitude
$x$. One example was to explore glueball decay to two pseudoscalar 
mesons in the quenched approximation~\cite{wein}.

A careful discussion of the matching~\cite{ll} between finite volume and
 infinite volume relies on a quantitative treatment of the interactions 
between the two bodies ($A+B$). In our treatment, we are neglecting this
 interaction, so one can obtain the matching directly from  phase space
considerations~\cite{cmdecay, wein}. The key step is that we are
normalising the  states $H$, $A$ and $B$ to one. Then the density of
$A+B$ states in energy is given from  $E({\bf n})= \sqrt{m_A^2 + k^2} +
\sqrt{m_B^2 + k^2}$ with ${\bf k}= 2 \pi {\bf n}/L$. In our application
here, we shall treat  $m_A$ as infinite so the density of states $
\rho(E)=4 \pi n^2 dn/dE = L^3 k E_B /(2 \pi^2)$. Then first order
perturbation theory (Fermi's Golden Rule) implies  a transition rate
$\Gamma=2 \pi x^2 \rho(E)$. Here we explicitly see the  factor of $L^3$
from the density of states cancelling the implicit factor of $L^{-3/2}$
in $x$. At first sight the fact that the density of states  is rather
sparse in a finite volume is worrying~\cite{testa} - but the matching is
 done at the level of the transition amplitude so the density of states
is  needed in a large volume only.

 In conclusion, we are able to evaluate the transition amplitude $x$ on
a lattice when  $E_H \approx E_A + E_B$ for momentum $k$ which is the
minimum lattice momentum at that finite volume, provided
 \be
   (E_H-E_A-E_B) t << 5, \ \    xt << 1,\ \ {\rm and} \ \ (E'-E)t >> 1
 \ee 
 where $E'-E$ is the energy gap to an excited state.
  Then we use this lattice result to determine the large volume physical
decay rate where  $k$ is the momentum of the decay product.

\section{Lattice transition matrix element}

\subsection{Energies}

 Here we use 207 dynamical fermion configurations~\cite{ukqcd} with 
$N_f=2$ flavours of sea-quark with masses around the strange quark mass
($\kappa$=0.1355 with NP improved clover having $C_{SW}=2.02$ at
$\beta=5.2$ for a 16$^3$ 32 lattice,  with lattice spacing given by
$r_0/a=5.04(4)$ which corresponds to $a$  of around 0.1 fm).

 We are interested in the heavy quark limit so we evaluate the usual 
static potential ($A_{1g}$) and the excited-gluonic potential which
corresponds to the $E_u$ representation. Results are shown in
fig.~\ref{pmvr}.

 We can now evaluate the energy release at each value of $R$ and compare
it  with known flavour-singlet meson masses. We are interested in the
minimum  momentum allowed which will be ${\bf n}=(1,0,0)$ for the scalar
meson emission  and ${\bf n}=(1,1,0)$ for the pseudoscalar emission. We
can evaluate the energies of these  states with non-zero momentum by
assuming the usual energy-momentum relationship  and taking the masses
as determined on a lattice~\cite{ukqcd}. We also check these energy
estimates from  our results here.   For example  from the pion mass
$am=0.294(4)$ and flavour singlet mass enhancement of around  0.06
one gets $aE(110)=0.66$ for this pseudoscalar state. For the scalar
meson  $am=0.628(30)$ implying aE(100)=0.74(3) for this scalar state. 
These two energy values are comparable with the energy release for $R$
values of around $0.2$ fm - see  fig.~\ref{pmvr} where this is
illustrated  for the scalar meson emission.

\subsection{Transitions}

We now discuss the transition matrix elements in  the heavy quark limit
where the quarkonia states will be accurately  treated by the static
approximation.  We then discuss the creation  of the light quark
antiquark pair which form the flavour singlet meson. 
 In particular we discuss how  to create operators for the $E_u$,
$A_{1g} + S(0^{++})$, and $A_{1g} + \eta$  states.

 Let the static quarks be separated by $R$ in the $z$-direction with the
 midpoint at ${\bf r}$. Then under rotations about the $z$-axis we 
have  a two-dimensional representation (like $J_z=1$). These two states 
correspond to flux  states from a lattice  operator which is the
difference of U-shaped paths from quark to antiquark of the form $\,
\sqcap - \sqcup$ where the transverse extent can be in the $x$ or $y$
direction respectively. 

 For the ground state ($A_{1g}$ on a lattice) we take a straight path
from  the static quark to antiquark. Then we need to discuss the spatial
distribution  of the light quark meson with respect to the static quarks. We
have to ensure that the  initial  and final states are in the same
representation of the symmetries  of the heavy quark state. This can be
achieved by constructing the two body state using the lattice operator

 \be
 O({\bf r}) = \sum_{\bf s} a({\bf r}) M({\bf s}) w({\bf s}, {\bf r}) 
 \ee
 where $a$ represents the colour field in the $z$-direction from ${\bf
r}-{\bf e_z}R/2$ to  ${\bf r}+{\bf e_z}R/2$ and $w$ is the distribution
function of the flavour singlet  meson operator $M$ (which will be
either a pseudoscalar or scalar meson). 
   Because of translational invariance, we can express this meson 
distribution function $w$ most efficiently in momentum space: 
 \be
   w({\bf s}, {\bf r}) = \sum_{\bf k} e^{i{\bf k}({\bf r-s})} w({\bf k})
  \ee

 The symmetries of $w({\bf r,s})$ depend on those of the meson produced.
 For scalar meson production (with $J^{PC}=0^{++}$), then $w({\bf s,r})$
is in an $E_u$ representation and this can be achieved by making $w({\bf
k})$ odd in $k_x$ and even in $k_y$ and $k_z$ where $x$ is the direction
of  transverse extent of the $E_u$ state described above. 

 For pseudoscalar meson production (which has $CP=-1$),   then $w({\bf
r,s})$ is in an $E_g$ representation and this can be achieved by making
$w({\bf k})$ odd in $k_y$ and $k_z$ and even in $k_x$. Another way to
see that this is the correct  symmetry configuration is from 
considering space inversions, since $P_x$, $P_y$ and $CP_z$ are 
conserved in the transition. Now consider the  $E_u$ representation
which is odd under $P_x$ and even under $P_ y$ and  $CP_z$, while the
$\eta$ operator, being pseudoscalar, is odd under  all three operations.
The $A_{1g}$ operator is even under all three operations, so we need to
introduce a wavefunction $w$  which is odd under inversions  $P_y$ and
$P_z$ (since $C=+1$ for the $\eta$) and even under $P_x$.

 In practice we evaluate the difference of two Wilson loops
corresponding to creating the $E_u$  state at ${\bf r},\ t$ with
transverse  extent in the $x$ and $-x$ directions and annihilating the
$A_{1g}$ state at ${\bf r},\ t+T$.  Let us call this observable
$\AE({\bf r})$ and its spatial Fourier transform
 $\AE({\bf q})$. The disconnected fermion loop (from operator
$\eta=\bar{q} \gamma_5 q$ for pseudoscalar mesons  or  $S=\bar{q}q$ for
scalar mesons)  is evaluated at each spatial point ${\bf s}$ at time
$t+T$ and its Fourier transform is $M({\bf p})$ corresponding to 
$\eta({\bf p})$ or $S({\bf p})$. Then the required correlation is given 
after summing over ${\bf r}$ as 

 \be
 \sum_{\bf k} w({\bf k}) \AE(-{\bf k}) M({\bf k})
 \ee

 Here $\AE$ represents a Wilson loop which has a zero expectation on its
own since it  has an $E_u$ state at one end and an $A_{1g}$ state at the
other end in time, and $M$ represents a fermionic disconnected loop with
non-zero momentum which also  has a zero expectation value on its own.
Here $M$ is evaluated by stochastic methods~\cite{cmcm}. The product of these
two operators is constructed to have  a non-zero expectation value and
that is the target of this investigation.
 We actually used fuzzed sources for the spatial ends of $\AE$  (2 or 13
iterations   of $U \to {\cal P} (c U_{\rm Straight}+ \sum U_{\rm
Staples})$ with  $c=2.5$ for the $A_{1g}$ end but only the higher
iteration level for  the $E_u$ end) and different  sizes (1 or 2 lattice
spacings) for the transverse extent of the $E_u$ end  while we use
fuzzed and local fermionic operators for the light-quark meson.

 We have here described the correlation in terms of a specific
orientation  of the static quark separation and of the transverse extent
of the $E_u$ state.  On a lattice we sum over all cubic rotations,
translations and reflections to  increase statistics.

\subsubsection{Pseudoscalar decays}

 For pseudoscalar decays, since $w({\bf k})$ is odd in both $k_y$ and in
$k_z$, these momenta must be non-zero. The simplest assumption which
corresponds to the lightest allowed state,  is used in this exploratory
study; namely, that  $n_x=0,\ n_y=\pm 1$ and $n_z=\pm 1$  where the
lattice momentum ${\bf k} =2 {\bf n} \pi/ L$. In terms of these
components of the momentum, the required  correlation is

 \be
 2\Re ( \AE(0,1,1) \eta(0,-1,-1) - \AE(0,1,-1) \eta(0,-1,1)
 \ee 

 which we evaluate as (here $cc$ means cos transform in $y$ and $z$, etc.).

 \be
  4(-\AE_{cc} \eta_{ss}-\AE_{ss} \eta_{cc}+
 \AE_{sc} \eta_{cs}+\AE_{cs} \eta_{sc})
 \ee

 Note that $\eta({\bf s})$ is real for Wilson-like fermion formalisms -
so we take the real part of the stochastic  estimate, while  $\AE({\bf r})$
is complex since the Wilson loop in SU(3) has an  orientation, but
$\eta$ has even charge conjugation so we  need to take the real part
here also.

 For this minimum momentum, we have energy equality at $R=2a$. 

 Following eq.~\ref{xrat}, we normalise states to 1 and evaluate the 
transition matrix element $x= \langle H | A \eta \rangle$  from the
ratio at each $t$ value 
 \be 
 xt = {C_{H-A\eta}(t)  \over(C_{A-A}(t)\, C_{H-H}(t)\, 
  C_{\eta-\eta}(t))^{1/2} }
 \label{etaratio}
 \ee
 where we have neglected interactions in the $A \eta$ two body state so
have  used for its correlator the direct product of the $A$ and $\eta$ 
propagation. This amounts to neglecting the correlation between the  $R
\times t$ Wilson loop giving $C_{A-A}(t)$ and the $\eta$ correlator
involving  light quarks, so that
 \be
   C_{A\eta-A\eta}(t) = C_{A-A}(t) C_{\eta-\eta}(t)
 \ee
 Here the $C_{\eta-\eta}(t)$ contribution includes the connected and
disconnected  contributions to the $\eta$ propagation.

 We obtain no signal for the ratio of eq.~\ref{etaratio} which curtails
our investigation. We can obtain limits, however. For instance at $R=2a$
and at $t=1$, a value $xt=0 \pm 0.0009$ is obtained. This can be turned
into a limit on this transition rate of $\Gamma < 1$ MeV. Note that this
value is for quarks of strange mass, at $R=0.2$ fm so with no account of
wavefunction effects, for $N_f=2$ with no account  of $\eta,\ \eta'$
mixing, with no account of excited state  contamination and without any
continuum limit. Because of the lack of any signal, we are unable to
pursue  these corrections and extrapolations. Perhaps the most useful
conclusion is that this  transition appears weak, maybe because it is a
D-wave and so involves cancellations between  different spatial
components of the $\eta$ wave function.

\subsubsection{Scalar decays}

 For scalar decays, since $w({\bf k})$ is odd in $k_x$,  this momentum
must be non-zero. The simplest assumption which corresponds to the
lightest allowed state,  is used in this exploratory study; namely, that
 $n_x=\pm1,\ n_y=0$ and $n_z=0$.  For this minimum momentum, we have
energy equality in the transition at $R=2a$ - see fig.~\ref{pmvr}. In
terms of these components of the momentum, the required  correlation is

 \be
 \Im ( \AE(1,0,0) S(-1,0,0) - \AE(-1,0,0) S(1,0,0))
 \ee 

 which we evaluate as (here $c$ means cos transform in $x$, etc.).

 \be
  2 \Re (-\AE_{c} S_{s}+\AE_{s} S_{c})
 \ee

\begin{figure}[ht] 
 \epsfxsize=10cm\epsfbox{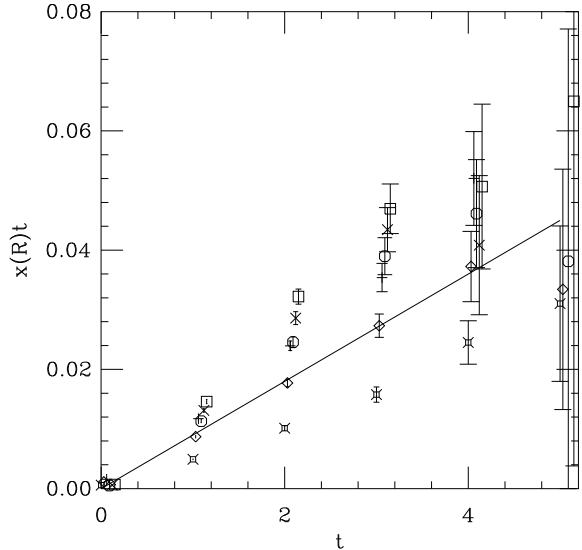}  

  \caption{ The transition matrix element $xt$  for $H \to A S$ with
momentum  ${\bf n}=(1,0,0)$ versus $t$. Here $R=0.1$ to 0.6 fm is
represented by symbols: fancy star, diamond, +, octagon, $\times$,
square. The line represents a linear  fit to the $R=0.2$ fm case.
 }
   \label{xfig}
  \end{figure}

We evaluate the  ratio at each $t$ value
 \be 
 xt = {C_{H-A S}(t)  \over(C_{A-A}(t) C_{H-H}(t) C_{S-S}(t))^{1/2} }
 \label{xscal}
 \ee
 where the scalar propagation in $C_{S-S}(t)$ again involves both connected
 and  disconnected contributions. In this case we do obtain a signal and
the values of $x$ extracted  are shown in fig.~\ref{xfig}.
 Moreover we do see good evidence for a linear dependence on $t$ as 
needed to ensure excited state contributions are removed. This linear 
dependence  sets in from very small $t$-values which may be explained 
if the off-diagonal transition matrix elements (ie the corresponding
$x$-values for the  transition from excited state to ground state) are
small compared to the diagonal case. From the slope we  can extract $x$
obtaining $ax = 0.009(1)$ at $R=2a$. This is indeed a small  value and
our assumptions about using the three point function analysis  are thus
fully justified. It would indeed be very difficult to detect directly
the shift of $\pm 0.01$  in the $aE$-values of fig.~\ref{pmvr} arising
from this mixing where the two levels cross at $R \approx 0.2$ fm.

Using this $x$-value and  an energy release of $aE=0.73$, we obtain a
transition rate  of $\Gamma=0.061(14)$ GeV. Note that this result is at
a fixed $R$-value (0.2 fm), for strange quarks and with no continuum
limit.

Although we are only able to evaluate $x$ at one $R$-value in principle
(where we have energy  equality) in this study, we can explore other $R$
values where the  energy equality is only approximate. Indeed since  the
energy difference increases to only $a\Delta \approx 0.35$ at $R=6/a$,
we find that the  criterion of approximate energy equality
(eq.~\ref{onshell}) is met for  the range of $t$-values considered here.
Moreover, we do find for $R$-values from $a$ to $6a$,  that the
correlator ratio of eq.~\ref{xscal} is consistent with linear in $t$ 
over the range of $t$ from $a$ to $5a$. Thus we can still estimate the
$x$-values by a linear fit in $t$, with  the understanding that excited
state effects may be less completely removed. We find an increase of $x$
with $R$ (see fig.~\ref{xfig}) with some sign of a saturation at large
$R$ (namely fit values of $ax=0.005(1),\ 0.009(1),\ 0.012(1),\
0.013(2),\ 0.015(2),\ 0.017(2)$ at $R=0.1,\dots 0.6$ fm, respectively).
 Since the transition to the scalar meson with momentum $k$ is a P-wave,
one would expect the  transition amplitude to have a factor of $k$.
However, we are working at a fixed volume,  so the minimum momentum $k$
is fixed as $R$ varies. Thus although the energy release varies with
$R$, we have a fixed momentum $k$ and hence this consideration should
not affect  the dependence of $x$ on $R$.

One way to interpret the $R$-dependence of $x$ is  by noting that the
scalar  meson wavefunction has a node at the centre of the $E_u$ state
in the transverse  direction (since $w$ is odd in relative transverse
spatial coordinate) and so it  is sensitive to the transverse width of
the excited gluonic flux in the  $E_u$ state. This 
increases~\cite{euflux}  with longitudinal  extent $R$ and then starts
to saturate, just as  we find.

\subsection{Phenomenology}

 In the extreme heavy quark limit, the heavy quarks are static and one
can define  a transition rate for each separation $R$. Also there will
be a well defined energy release for each value of $R$: for  example at
$R=2a=0.2$ fm we find  $E_{E_u}-E_{A_{1g}}= 0.73(1)/a= 1.4$ GeV. The 
energy of the scalar meson with the required momentum we take as 
$aE(100)=0.74(3)$ and we  also check that this energy  is consistent
with the value we find directly from fitting our scalar correlators with
this momentum (namely $aE(100) \approx 0.7$).  Thus indeed we  are close
to on-shell as required. In the real world, the quarks are  bound and
there is a distribution of $R$-values as given by the wave functions. 
For $b$ quarks,  as discussed previously, the static potentials allow us
to estimate the  quark wave functions. The hybrid wave function is
effectively P-wave and actually has  a quite large overlap with the
$\chi_b$ wave function (the overlap peaks at around 0.4 fm and the
wave-function overlap integrated over all $R$ gives a factor of 0.63  in
the transition rate assuming $x$ is independent of $R$).  The energy
release from the hybrid meson~\cite{hf8}  at 10.76(7) GeV to the
$\chi_b$  state at 9.893 GeV will be 0.87 GeV which is similar to  the
 value at fixed $R$ with $R \approx 0.4$ fm. Thus there is a mismatch in
the energy release we study  on the lattice (1.4 GeV) and that in
experiment (0.9 GeV).
 Note that this issue could be  very important:  the decay rate will be
proportional to $k^3$, so a small change of energy  release will have a
big effect on $k$ and an even bigger effect on the rate. Put more
bluntly: there will be no decay to  a scalar meson heavier than 870 MeV
in practice, but our estimates for the scalar  meson mass are indeed
heavier than this. Thus we will need some method to treat the virtual
(below threshold) production  of a scalar meson which subsequently
decays to two pions. We discuss this in the context of the quark mass 
dependence.

 We are evaluating the transition matrix element for quarks of mass
about  strange. At this quark mass on our lattice the scalar meson (mass
$ma=0.63$) is already  unstable to decay to two pions (mass $ma=0.29$
each) in an S-wave. Thus we  should consider, in principle, a further
layer of sophistication: the  sequential decay of the scalar meson to
two pions. This decay will become even  more significant as the quark
mass  is reduced further towards the physical case. Moreover it will
allow  scalar $\pi + \pi$ states to be produced even if the available
energy is less than the mass of a scalar meson as discussed above. Thus
we do not attempt a naive  extrapolation to  light quarks of realistic
mass. A study of the  three body final state will be needed to resolve
this issue more  completely.

 We could also study, in principle,  the transition for higher momentum,
for example $S$ having momentum $(1,1,0)2 \pi/L$  with energy
$aE(110)=0.85(4)$,  but then energy equality between initial and final
states would be at even smaller  $R$-values. Also  this higher energy
state would be coupled to the lighter $aE(100)$ state we  have explored
above and this would make the extraction from the lattice  more prone to
systematic errors. 

 Another possible avenue would be to vary the lattice spatial size $L$.
This  is not feasible with our current dynamical data set but is of
interest for the future.

 Without making a detailed study of the sequential decay of the scalar 
meson to two pions, our estimates of the decay rate will be qualitative.
 We find a rate of 61(14) MeV for the unphysical case of  a transition 
at $R=0.2$ fm  with strange quarks in the scalar meson. At the more
realistic value (where the wave function overlap peaks) of $R=0.4$ fm, 
we have a larger transition amplitude $ax=0.013(2)$ but the energy
release  is insufficient for decay to an on-shell scalar meson.
Including the  wavefunction overlap factor of 0.63, we conclude that the
decay rate to a scalar channel will  be less than 80 MeV (here the
upper limit  is from assuming that the  scalar meson is produced
off-shell with  momentum $k=2 \pi /L$ at R$ \approx 0.4$ fm).

 Flux-tube models have been used to estimate hybrid decay
widths~\cite{qmhyb}. For $b \bar{b}$ hybrid mesons,  they only consider
decays to  $BB$ and $B^* B$ and these decay rates are found to  be very
small (less than 1 MeV).

\section{Conclusions}

 We have presented arguments that, in the heavy quark limit, the decay
of  a $J^{PC}=1^{-+}$ spin-exotic hybrid meson will be primarily through
 flavour singlet light quark-antiquark emission.

 We have explored in lattice QCD with $N_f=2$ flavours of sea quark
(with mass near that of the  strange quark) the transition between an
excited gluonic state with heavy quarks at separation $R$  and a ground
state gluonic system with a flavour-singlet light quark-antiquark pair 
emitted. We find a very weak transition amplitude for emission of a
pseudoscalar meson but a much larger  rate for a scalar meson.

 On a lattice it is only possible to extract a limited set of
information: namely  when there is an on-shell transition in the  finite
volume used.  Our raw lattice results are a transition width of less
than 1 MeV for the pseudoscalar case  and 61(14) MeV for the scalar
case. These results are for transitions  at fixed $R \approx 0.2$ fm in
the heavy quark limit, for light quarks that are of strange mass.

 For $b$ quarks the relevant transitions will be $H \to \chi_b \eta$ and
 $H \to \chi_b S$. We argue that  wave function effects will  suppress
these decay rates rather little (a factor of 0.6) while the choice of a 
more appropriate $R$-value (of 0.4 fm) will increase the rates. This
yields  an off-shell decay rate to a scalar meson of around 80 MeV
which can be regarded as an upper limit.  The main uncertainty comes
from the sensitive  dependence (like $k^3$) of the rate on the energy
release and the complications caused by  the subsequent decay of the
scalar meson to two pions. More work needs to be done to build 
phenomenological models of scalar meson production and decay and,
eventually,  to explore the transition to two pions directly on the 
lattice.

Despite  this, we consider that first principles QCD evaluation of 
these hadronic transitions is a very valuable component of a
phenomenological  study of hybrid decays. Our  results are consistent
with the expectation that these spin-exotic hybrid meson states  are
relatively narrow and hence will be detectable experimentally. 
  We have not evaluated decay processes that are not allowed in the
heavy quark limit (such as retardation effects or  heavy quark
spin-flip) and it  would be valuable to investigate them to ensure that
they are indeed negligible compared to the string de-exitation decay that 
we find to be important.

\end{document}